\documentclass[aps,preprint]{revtex4}%
\usepackage{amsfonts}
\usepackage{amsmath}
\usepackage{amssymb}
\usepackage{graphicx}%
\begin{document}
\title{{\LARGE SHEAR AND VORTICITY IN INFLATIONARY BRANS-DICKE COSMOLOGY WITH
LAMBDA-TERM}}
\author{Marcelo Samuel Berman$^{1}$}
\affiliation{$^{1}$Instituto Albert Einstein\ - Av. Candido Hartmann, 575 - \ \# 17}
\affiliation{80730-440 - Curitiba - PR - Brazil - E-mail: msberman@institutoalberteinstein.org}
\begin{abstract}
We find a solution for exponential inflation in Brans-Dicke cosmology endowed
with a cosmological term, which includes time-varying shear and vorticity. We
find that the scalar field and the scale factor increase exponentialy while
shear, vorticity, energy density, cosmic pressure and the cosmological term
decay exponentialy for beta
$<$
0, where beta is defined in the text.

\end{abstract}
\maketitle

\begin{center}
{\LARGE SHEAR AND VORTICITY IN INFLATIONARY BRANS-DICKE COSMOLOGY WITH
LAMBDA-TERM}

\bigskip

Marcelo Samuel Berman
\end{center}

\bigskip

\bigskip It appears that the existence of shear ( $\sigma$\ ), vorticity (
$\varpi$\ ) and cosmological "constant" ( $\Lambda$\ ), have not been well
discussed in the context of a scalar field theory, say, compatible with
Brans-Dicke (1961) theory.

\bigskip

Not only for scalar-tensor theory, as such, but also for string researchers,
the introduction of non-null shear, vorticity and lambda, for inflationary
models, stand in the goals of theorists.

\bigskip

Though initially masterminded in order to fulfill Machian ideas, as conceived
by Einstein, it has revealed much more fruitful for many reasons, including
the possibility of extension to a time-varying increasing coupling "constant",
which would make gravity theory (read "Scalar-Tensor"), indistinguishable from
General Relativity, in matter-dominated epochs. The origin of all scalar
tensor theories rests, in the original Brans-Dicke (B.D.) theory.

\bigskip

Conventional B.D. theory is stated in the "Jordan frame", i.e., it conforms
with the original paper (Brans and Dicke, 1961). Another picture is called the
"Einstein's frame", or non-conventional system of units; the latter, resembles
as much as possible, with General Relativity, but is unable to account for
vacuum Physics, when Einstein's tensor is null. The reason is the
\textit{omni} -\ present scalar field, which provides for non-vacuum state.
The scalar field is a cause for gravity, so it can not be removed. Covariant
divergence conservation equation, \ is not obeyed, as well as the usual
general relativistic equations for geodesics, and geodesic deviation.

It is generally accepted that scalar tensor cosmologies play a central
r\^{o}le in the present view of the very early Universe (Faraoni, 2004). The
cosmological "constant" (Berman, 2007), which represents quintessence, is a
time varying entity, whose origin remounts to Quantum theory. The first, and
most important scalar tensor theory was devised by Brans and Dicke(1961).
Afterwards, Dicke(1962) presented a new version of the theory, where the field
equations resembled Einstein's equations, but time, length, and inverse mass,
were scaled by a factor \ $\phi^{-\frac{1}{2}}$\ \ where \ $\phi$\ \ stands
for the scalar field: it is the Einstein's frame units framework. Then, the
energy momentum tensor \ \ $T_{ij}$\ \ is augmented\ by a new term
$\Lambda_{ij}$\ , so that:

\bigskip

$G_{ij}=-8\pi G\left(  T_{ij}+\Lambda_{ij}\right)  $\ \ \ \ \ \ \ \ \ \ , \ \ \ \ \ \ \ \ \ \ \ \ \ \ \ \ \ \ \ \ \ \ \ \ \ \ \ \ \ \ \ \ \ \ \ \ \ \ \ \ \ \ \ \ \ \ \ \ \ \ \ \ \ \ \ \ \ (1)

\bigskip

where \ \ $G_{ij}$\ \ stands for Einstein's tensor. The new energy tensor
quantity, including a \ $\Lambda g_{\mu\nu}$\ \ term, is given by new modified
energy density, with added \ $\frac{\Lambda}{\kappa}$\ \ term, while cosmic
pressure is subtracted by the same \ \ $\frac{\Lambda}{\kappa}$\ \ term; we
have, then,

\bigskip

$\Lambda_{ij}=\frac{2\omega+3}{16\pi G\phi^{2}}\left[  \phi_{i}\phi_{j}%
-\frac{1}{2}G_{ij}\phi_{k}\phi^{k}\right]  $ \ \ \ \ \ \ \ \ \ \ \ \ . \ \ \ \ \ \ \ \ \ \ \ \ \ \ \ \ \ \ \ \ \ \ \ \ \ \ \ \ \ \ \ \ \ \ \ \ \ \ \ \ \ (2)

\bigskip

In the above, \ $\omega$\ \ is the coupling constant. The other equation is:

\bigskip

$\square\log\phi=\frac{8\pi G}{2\omega+3}T$\ \ \ \ \ \ , \ \ \ \ \ \ \ \ \ \ \ \ \ \ \ \ \ \ \ \ \ \ \ \ \ \ \ \ \ \ \ \ \ \ \ \ \ \ \ \ \ \ \ \ \ \ \ \ \ \ \ \ \ \ \ \ \ \ \ \ \ \ \ \ \ \ \ \ \ \ (3)

\bigskip

where \ $\square$\ \ is the generalized d'Alembertian, and $T=T_{i}^{i}%
$\ \ \ .\ \ It is useful to remember that the energy tensor masses are also
scaled by \ $\phi^{-\frac{1}{2}}$\ \ .

\bigskip

For\ the Robertson-Walker's flat metric,

\bigskip

$ds^{2}=dt^{2}-\frac{R^{2}(t)}{\left[  1+\left(  \frac{kr^{2}}{4}\right)
\right]  ^{2}}d\sigma^{2}$ \ \ \ \ \ \ \ \ \ \ \ \ \ \ , \ \ \ \ \ \ \ \ \ \ \ \ \ \ \ \ \ \ \ \ \ \ \ \ \ \ \ \ \ \ \ \ \ \ \ \ \ \ \ \ \ \ \ \ \ (4)

\bigskip

where \ \ $k=0$\ \ and \ $d\sigma^{2}=dx^{2}+dy^{2}+dz^{2}$\ \ .

\bigskip

The field equations now read, in the alternative Brans-Dicke
reformulation(Raychaudhuri, 1979), for a perfect fluid,

\bigskip

$\frac{8\pi G}{3}\left(  \rho+\frac{\Lambda}{\kappa}+\rho_{\lambda}\right)
=H^{2}\equiv\left(  \frac{\dot{R}}{R}\right)  ^{2}$ \ \ \ \ \ \ \ \ \ \ \ . \ \ \ \ \ \ \ \ \ \ \ \ \ \ \ \ \ \ \ \ \ \ \ \ \ \ \ \ \ \ \ \ \ \ \ \ \ \ (5)

\bigskip

$-8\pi G\left(  p-\frac{\Lambda}{\kappa}+\rho_{\lambda}\right)  =H^{2}%
+\frac{2\ddot{R}}{R}$ \ \ \ \ \ \ \ \ \ \ \ \ \ \ \ \ . \ \ \ \ \ \ \ \ \ \ \ \ \ \ \ \ \ \ \ \ \ \ \ \ \ \ \ \ \ \ \ \ \ \ \ \ \ (6)

\bigskip

In the above, we have: \ 

$\bigskip$

$\rho_{\lambda}=\frac{2\omega+3}{32\pi G}\left(  \frac{\dot{\phi}}{\phi
}\right)  ^{2}=\rho_{\lambda0}\left(  \frac{\dot{\phi}}{\phi}\right)  ^{2}$
\ \ \ \ \ \ \ . \ \ \ \ \ \ \ \ \ \ \ \ \ \ \ \ \ \ \ \ \ \ \ \ \ \ \ \ \ \ \ \ \ \ \ \ \ \ \ \ \ \ \ \ \ \ \ (7)

\bigskip

\bigskip The complete set of field equations, must be complemented by three
more equations, of which only two are independent, when taken along with
\ (5),(6) and (7): one is the dynamical fluid equation; the other is the
continuity one, as follow:

\bigskip

$\frac{d}{dt}\left[  \left(  \rho+\rho_{\lambda}+\frac{\Lambda}{\kappa
}\right)  R^{3}\right]  +3R^{2}\dot{R}\left[  p-\frac{\Lambda}{\kappa}%
+\rho_{\lambda}\right]  =0$\ \ \ \ \ \ \ \ , \ \ \ \ \ \ \ \ \ \ \ \ \ \ \ \ \ \ \ \ \ \ \ (8a)

\bigskip

and,

\bigskip

$\frac{d}{dt}\left[  \left(  \rho+\frac{\Lambda}{\kappa}\right)  R^{3}\right]
+3R^{2}\dot{R}\left[  p-\frac{\Lambda}{\kappa}\right]  +\frac{1}{2}R^{3}%
\frac{\dot{\phi}}{\phi}\left[  \rho+\frac{4\Lambda}{\kappa}-3p\right]
=0$\ \ \ \ \ \ . \ \ \ \ \ (8b)

\bigskip

All the above equations, are generalizations of those in the book by
Raychaudhuri (1979), which were written for \ $\Lambda=0$\ . We stress again,
that the equations refer to a perfect fluid. When we combine the above
equations, we would find equation (8) below:\ 

\bigskip

$\frac{\ddot{R}}{R}=-\frac{4\pi G}{3}\left(  \rho+3p+4\rho_{\lambda}%
-\frac{\Lambda}{4\pi G}\right)  $ \ \ \ \ \ \ \ \ \ \ \ \ \ . \ \ \ \ \ \ \ \ \ \ \ \ \ \ \ \ \ \ \ \ \ \ \ \ \ \ \ \ \ \ \ \ \ \ \ \ \ (8)

\bigskip

\bigskip Relation (8) represents Raychaudhuri's equation for a perfect fluid.
By the usual procedure, we would find the Raychaudhuri's equation in the
general case, involving shear ($\sigma_{ij}$) and vorticity ($\varpi_{ij}$);
the acceleration of the fluid is null for the present case.

\bigskip

\bigskip It suffices to make the following substitution in Raychaudhuri's equation:

\bigskip

$\left[  -4\pi G\left(  \rho+3p+4\rho_{\lambda}\right)  \right]
\rightarrow\left[  -4\pi G\left(  \rho+3p+4\rho_{\lambda}\right)  +2\left(
\varpi^{2}-\sigma^{2}\right)  \right]  $ \ \ \ ,

and then we find:

\bigskip

$3\dot{H}+3H^{2}=2\left(  \varpi^{2}-\sigma^{2}\right)  -4\pi G\left(
\rho+3p+4\rho_{\lambda}\right)  +\Lambda$ \ \ \ \ \ \ \ \ , \ \ \ \ \ \ \ \ \ \ \ \ \ \ (9)\ \ \ 

\bigskip

where \ $\Lambda$\ \ stands for a cosmological "constant". As we are mimicking
Einstein's field equations, $\Lambda$\ \ in (9) stands like it were a constant
(see however, for instance, Berman, 2007). Notice that, \ when we impose that
the fluid is not accelerating, this means that the quadri-velocity is tangent
to the geodesics, i.e., the only interaction is gravitational. The trick
leading to equation (9), from equation (8), may not have been considered
explicitly \ elsewhere, like we do here.

\bigskip

Now, we remember that, if there were no other important reason to consider
inflation, we would still mention the fact that, \ with a Machian equation of
state, as considered by Berman (2006), we would have a graceful entrance into
exponentially growing scalar-factor. (Berman, \ 2006).

\bigskip

So, consider now exponential inflation, like we find in Einstein's theory:

\bigskip

$R=R_{0}e^{Ht}$\ \ \ \ \ \ \ \ , \ \ \ \ \ \ \ \ \ \ \ \ \ \ \ \ \ \ \ \ \ \ \ \ \ \ \ \ \ \ \ \ \ \ \ \ \ \ \ \ \ \ \ \ \ \ \ \ \ \ \ \ \ \ \ \ \ \ \ \ \ \ \ \ \ \ \ \ \ \ \ \ \ \ \ \ (10)

\bigskip

and,

\bigskip

$\Lambda=3H^{2}$ \ \ \ \ \ \ \ \ \ \ .

\bigskip

For the time being, $H$ \ is just a constant, defined by \ $H=\frac{\dot{R}%
}{R}$\ \ . We shall see, when we go back to conventional\ \ Brans-Dicke
theory, that \ $H$\ \ is not the Hubble's constant.

\bigskip

From (10), we find $H=H_{0}=$\ constant.\ 

\bigskip

A\ solution of Raychaudhuri's equation\ (9), together with all other
equations, would be the following:

\bigskip

$\varpi=\varpi_{0}e^{-\frac{\beta}{2}t}$ \ \ \ ; \ \ \ \ 

\bigskip

$\sigma=\sigma_{0}e^{-\frac{\beta}{2}t}$ \ \ \ ;

\bigskip

$\rho=\rho_{0}e^{-\beta t}$ \ \ \ ;

\bigskip

$p=p_{0}e^{-\beta t}$ \ \ \ ; \ \ \ \ \ \ \ \ \ \ \ \ \ \ \ \ \ \ \ \ \ \ \ \ \ \ \ \ \ \ \ \ \ \ \ \ \ \ \ \ \ \ \ \ \ \ \ \ \ \ \ \ \ \ \ \ \ \ \ \ \ \ \ \ \ \ \ \ \ \ \ \ \ \ \ \ \ \ \ \ \ (11)

\bigskip

$\Lambda=\Lambda_{0}=$ constant.

\bigskip

\bigskip In the above, \ $\sigma_{0}$\ , \ $\varpi_{0}$\ \ , \ \ $p_{0}$\ \ ,
\ \ $\rho_{0}$\ \ and \ \ $\beta$\ \ are constants, obeying conditions (11b)
and (11c) below:

\bigskip

$\varpi_{0}^{2}-\sigma_{0}^{2}=2\pi G\left[  \rho_{0}+3p_{0}+4\rho_{\lambda
0}\right]  $ \ \ \ \ \ \ \ , \ \ \ \ \ \ \ \ \ \ \ \ \ \ \ \ \ \ \ \ \ \ \ \ \ \ \ \ \ \ \ \ \ \ \ \ \ \ \ \ \ \ (11b)

\bigskip

and,

\bigskip

$\Lambda_{0}=3H_{0}^{2}$ \ \ \ \ \ \ \ \ \ \ \ . \ \ \ \ \ \ \ \ \ \ \ \ \ \ \ \ \ \ \ \ \ \ \ \ \ \ \ \ \ \ \ \ \ \ \ \ \ \ \ \ \ \ \ \ \ \ \ \ \ \ \ \ \ \ \ \ \ \ \ \ \ \ \ \ \ \ \ \ \ \ \ \ \ \ (11c)

\ \ 

\bigskip\bigskip The ultimate justification for this solution is that one
finds a good solution in the conventional units theory. It can be check easily
that, when the perfect fluid part of all equations produced above, are
fulfilled, then (9) gives the imperfect fluid Raychaudhuri's equation. In
other words, \ $\rho$\ \ and \ \ $p$\ \ \ are the usual energy density and
cosmic pressure terms.\ 

\bigskip

When we return to conventional units, we retrieve the following corresponding solution:

\bigskip

$\bar{R}=R_{0}\phi^{\frac{1}{2}}e^{Ht}$ \ \ \ \ \ \ \ \ ; \ \ \ \ \ \ \ \ \ \ \ \ \ \ \ \ \ \ \ \ \ \ \ \ \ \ \ \ \ \ \ \ \ \ \ \ \ \ \ \ \ \ \ \ \ \ \ \ \ \ \ \ \ \ \ 

\bigskip

$\bar{\rho}=\rho_{0}\phi^{-2}e^{-\beta t}$\ \ \ \ \ \ \ \ \ ;

\bigskip

$\bar{p}=p_{0}\phi^{-2}e^{-\beta t}=\left[  \frac{p_{0}}{\rho_{0}}\right]
\bar{\rho}$\ \ \ \ \ \ \ \ \ ;

\bigskip\ \ \ \ \ \ \ \ \ \ \ \ \ \ \ \ \ \ \ \ \ \ \ \ \ \ \ \ \ \ \ \ \ \ \ \ \ \ \ \ \ \ \ \ \ \ \ \ \ \ \ \ \ \ \ \ \ \ \ \ \ \ \ \ \ \ \ \ \ \ \ \ \ \ \ \ \ \ \ \ \ \ \ \ \ \ \ \ \ \ \ \ \ \ \ \ \ \ \ \ (12)

$\bar{\varpi}=\varpi\phi^{-\frac{1}{2}}$\ \ \ \ \ \ \ \ \ \ \ \ \ \ ;

\bigskip

$\bar{\sigma}=\sigma\phi^{-\frac{1}{2}}$ \ \ \ \ \ \ \ \ \ \ \ \ \ \ ;

\bigskip

$\bar{\Lambda}=\Lambda_{0}\phi^{-1}$\ \ \ \ \ \ \ \ \ \ \ \ \ ;

\bigskip

\bigskip$\bar{\phi}=\phi=\phi_{0}e^{-\frac{\beta}{2}\sqrt{A}\text{
\ }e^{-\frac{\beta}{2}t}}$ \ \ \ \ \ \ \ \ \ ;

\bigskip

$\bar{H}=H\phi^{-\frac{1}{2}}$ \ \ \ \ \ \ \ \ \ \ \ \ \ \ \ \ \ \ \ \ \ \ \ .

As we promised to the reader, $H$\ is not the Hubble's constant. Instead, we find:

\bigskip

$\bar{H}^{2}=\frac{1}{3}\bar{\Lambda}$\ \ \ \ \ \ \ \ \ \ \ \ . \ \ \ \ \ \ \ \ \ \ \ \ \ \ \ \ \ \ \ \ \ \ \ \ \ \ \ \ \ \ \ \ \ \ \ \ \ \ \ \ \ \ \ \ \ \ \ \ \ \ \ \ \ \ \ \ \ \ \ \ \ \ \ \ \ \ \ \ \ \ \ \ (13)

\bigskip

\bigskip$\bar{\Lambda}=\Lambda_{0}$ $\phi_{0}^{-1}$ $e^{\frac{\beta}{2}%
\sqrt{A}\text{ }e^{-\frac{\beta}{2}t}}$ \ \ \ \ ; \ \ \ \ \ \ \ \ \ \ \ \ \ \ \ \ \ \ \ \ \ \ \ \ \ \ \ \ \ \ \ \ \ \ \ \ \ \ \ \ \ \ \ \ \ \ \ \ \ \ \ \ \ \ \ \ \ \ \ \ \ (14)

\bigskip

$\bar{\rho}=\rho_{0}$ $\phi_{0}^{-2}$ $e^{\beta\left[  \sqrt{A}\text{
\ }e^{-\frac{\beta}{2}\text{ }t}-\text{ }t\right]  }$ \ \ \ \ \ \ ;\ \ \ \ \ \ \ \ \ \ \ \ \ \ \ \ \ \ \ \ \ \ \ \ \ \ \ \ \ \ \ \ \ \ \ \ \ \ \ \ \ \ \ \ \ \ \ \ \ \ \ \ \ \ (15)

\bigskip

$\bar{p}=p_{0\text{ }}\phi_{0}^{-2}$ $e^{\beta\left[  \sqrt{A}\text{
\ }e^{-\frac{\beta}{2}\text{ }t}-\text{ }t\right]  }$ \ \ \ \ \ \ ;\ \ \ \ \ \ \ \ \ \ \ \ \ \ \ \ \ \ \ \ \ \ \ \ \ \ \ \ \ \ \ \ \ \ \ \ \ \ \ \ \ \ \ \ \ \ \ \ \ \ \ \ \ \ (16)

\bigskip

$\bar{R}=R_{0}$ $\phi_{0}^{-\frac{1}{2}}$ $e^{\left[  H\text{ }t\text{ }%
-\frac{1}{4}\beta\text{ }\sqrt{A}\text{ \ }e^{-\frac{\beta}{2}\text{ }%
t}\right]  }$ \ \ \ \ \ \ ;\ \ \ \ \ \ \ \ \ \ \ \ \ \ \ \ \ \ \ \ \ \ \ \ \ \ \ \ \ \ \ \ \ \ \ \ \ \ \ \ \ \ \ \ \ (17)

\bigskip

\bigskip$\bar{\varpi}=\varpi_{0}$ $\phi_{0}^{-\frac{1}{2}}$ $e^{-\frac{1}%
{2}\beta\left[  \text{ }t\text{ }-\frac{1}{2}\text{ }\sqrt{A}\text{
\ }e^{-\frac{\beta}{2}\text{ }t}\right]  }$ \ \ \ \ \ \ ;\ \ \ \ \ \ \ \ \ \ \ \ \ \ \ \ \ \ \ \ \ \ \ \ \ \ \ \ \ \ \ \ \ \ \ \ \ \ \ \ \ \ (18)

\bigskip

\bigskip$\bar{\sigma}=\sigma_{0}$ $\phi_{0}^{-\frac{1}{2}}$ $e^{-\frac{1}%
{2}\beta\left[  \text{ }t\text{ }-\frac{1}{2}\text{ }\sqrt{A}\text{
\ }e^{-\frac{\beta}{2}\text{ }t}\right]  }$ \ \ \ \ \ \ ,\ \ \ \ \ \ \ \ \ \ \ \ \ \ \ \ \ \ \ \ \ \ \ \ \ \ \ \ \ \ \ \ \ \ \ \ \ \ \ \ \ \ \ \ (19)

\bigskip

and,

\bigskip

$\bar{H}=H$ $\phi_{0}^{-\frac{1}{2}}$ $e^{\frac{1}{4}\beta\sqrt{A}\text{
\ }e^{-\frac{\beta}{2}\text{ }t}}$ \ \ \ \ \ \ .\ \ \ \ \ \ \ \ \ \ \ \ \ \ \ \ \ \ \ \ \ \ \ \ \ \ \ \ \ \ \ \ \ \ \ \ \ \ \ \ \ \ \ \ \ \ \ \ \ \ \ \ \ \ \ \ (20)

From the hypothesis of an expanding Universe, we have to impose:

\bigskip

$\bar{H}>0$ \ \ \ \ \ . \ \ \ \ \ \ \ \ \ \ \ \ \ \ \ \ \ \ \ \ \ \ \ \ \ \ \ \ \ \ \ \ \ \ \ \ \ \ \ \ \ \ \ \ \ \ \ \ \ \ \ \ \ \ \ \ \ \ \ \ \ \ \ \ \ \ \ \ \ \ \ \ \ \ \ \ \ \ \ \ \ (21)

\bigskip

\bigskip We can not forget that we still have the conditions (11b) and (11c)
to be obeyed by the constants.

\bigskip

\bigskip We now investigate the limit when \ \ $t\longrightarrow\infty$\ \ of
the above formulae, having in mind that, by checking that limit, \ we will
know which ones increase or decrease with time; of course, we can not stand
with an inflationary period unless it takes only an extremely small period of
time. We shall suppose that \ $\beta<0$\ \ .

\bigskip

We find:

\bigskip

$\lim\limits_{t\longrightarrow\infty}\bar{H}=0$ \ \ \ \ ;\ \ \ 

\bigskip

$\lim\limits_{t\longrightarrow\infty}\bar{R}=\infty$ \ \ \ \ ;

\bigskip

$\lim\limits_{t\longrightarrow\infty}\bar{\sigma}=\lim
\limits_{t\longrightarrow\infty}\bar{\omega}=0$ \ \ \ \ ;

\bigskip

$\lim\limits_{t\longrightarrow\infty}\bar{\rho}=\lim\limits_{t\longrightarrow
\infty}\bar{p}=0$ \ \ \ \ \ \ ;

\bigskip

$\lim\limits_{t\longrightarrow\infty}\bar{\Lambda}=0$ \ \ \ \ ;

\bigskip

$\lim\limits_{t\longrightarrow\infty}\bar{\phi}=\infty$ \ \ \ \ .

\bigskip

\bigskip\bigskip As we can check, the scale factor and the scalar field are
time-increasing, while all other elements of the model, namely, vorticity,
shear, Hubble's parameter, energy density, cosmic pressure, and cosmological
term, as described by the above relations, decay with time. This being the
case, shear and vorticity decaying, the tendency is that, after inflation, we
retrieve\ \ a nearly perfect fluid: \ inflation has the peculiarity of
removing shear and vorticity from the model. It has to be remarked, that
pressure and energy density obey a perfect gas equation of state. This is the
outcome for the \textit{"graceful" exit \ }from the inflationary phase, while
we adopt into consideration, for instance, \ the section with the latter title
(in italics), of the book by Kolb and Turner (1990), to which we refer the reader.

\bigskip\bigskip

{\Large Acknowledgements}

\bigskip

I, gratefully, thank an anonymous referee, and my intellectual mentors,
Fernando de Mello Gomide and M. M. Som, and I am also grateful for the
encouragement by Albert, Paula and Geni. Further thanks go to Nelson Suga,
Antonio F. da F. Teixeira, Marcelo Fermann Guimar\~{a}es and Mauro Tonasse.

\bigskip

\bigskip

{\Large References}

\bigskip

\bigskip Berman,M.S. (2006) - Chapter 5 of \textit{New Developments in Black
Hole Research, }edited by Paul Kreitler, Nova Science, New York; Chapter 5, of
\textit{Trends in Black Hole Research, }edited by Paul Kreitler, Nova Science,
New York.

Berman,M.S. (2007) - \textit{Introduction to General Relativity and the
Cosmological Constant Problem}, Nova Science, New York. [just published!!!]

\bigskip Brans, C.; Dicke, R.H. (1961) - Physical Review, \textbf{124}, 925.

Dicke, R.H. (1962) - Physical Review, \textbf{125}, 2163.

Faraoni, V. (2004) - \textit{Cosmology in Scalar-Tensor Gravity, }Kluwer, Dordrecht.

Kolb, E.W.; Turner, M.S. (1990) - \textit{The early Universe, }Addison-Wesley,
Redwood City.

\bigskip Raychaudhuri, A. K. (1979) - \textit{Theoretical Cosmology, }Oxford
University Press, Oxford.

\bigskip
\end{document}